# Calculating the Expected Value of Sample Information using Efficient Nested Monte Carlo: A Tutorial

Anna Heath, Gianluca Baio

## 1. Introduction

The Expected Value of Sample Information (EVSI) [1] uses evidence about the cost and effectiveness of new treatments to determine the expected economic benefit of undertaking a proposed study [2]. The EVSI calculates this value by determining the extent to which the additional information from the study reduces the probability and expected loss of making inefficient treatment recommendations [3]. In general, an inefficient recommendation would spend health resources that would be used to improve patient outcomes elsewhere. Additionally, information from future studies has the potential to improve patient outcomes within the disease area under investigation.

As the EVSI calculates the value of a specific trial, it has the potential to be used to determine the optimal allocation of research funding. Despite this, the EVSI has rarely been used in practical scenarios [4]. This is partly due to the large computational effort required to calculate the EVSI using the "gold-standard" nested Monte Carlo method [2].

Recently, a number of methods have been developed to reduce the computation time for the EVSI [5, 6, 7, 8, 9, 10]. While computationally efficient, these methods require the use of additional statistical techniques, e.g. specifying *sufficient statistics* [11], and may impose restrictions on the underlying economic model. Therefore, EVSI calculations have been repeatedly undertaken using Monte Carlo procedures, either using approximations to avoid nested simulations [12, 13, 14], or with high computational cost [15, 16, 17, 18]

Using nested Monte Carlo methods is advantageous as it can be used irrespective of the complexity of the economic model. Additionally, the EVSI is often understood in terms of this nested Monte Carlo estimation procedure [19, 20, 5, 8] making the computation method easier to comprehend. To utilise these advantages, Heath et al developed a computationally efficient EVSI calculation method based solely on nested sampling [10] to calculate the EVSI in all practical scenarios, within a reasonable timeframe.

The validity of this method has been demonstrated elsewhere [10] and thus this paper is focused on presenting its practical implementation. Section 2 introduces the EVSI, key notation and the nested Monte Carlo method. Section 3 outlines the efficient Monte Carlo method using a simple "toy". Finally, Section 4 investigates the optimal choice of the key input for this efficient Monte Carlo method using an economic model evaluating a new Chemotherapy treatment [21].

## 2. Formal definition of the EVSI

In general, information has value as it reduces uncertainty in the inputs of a health economic model, denoted $\boldsymbol{\theta}$. Therefore, to calculate the EVSI, we must model the current level of uncertainty in $\boldsymbol{\theta}$ using probability distributions that indicate the most plausible range for these inputs. This is performed as part of a Probabilistic Sensitivity Analysis (PSA) [22, 23] where the parameters of these distributions are usually informed by literature reviews alongside clinical trials and expert opinion.

To value each of the treatment options, normally using a net benefit function [24], the model inputs are combined using an economic model. To ease our explanation, we assume throughout that this economic model only compares two treatment options – typically, an innovative treatment and the standard of care. The output of the economic model is then the incremental net benefit (INB), i.e. the difference between the net benefit of the innovative treatment and net benefit of the standard of care. In this setting, uncertainty in the model inputs implies that the INB is also uncertain. From this, we conclude that the innovative treatment is optimal given current uncertainty if the average INB is greater than 0 and the standard of care is optimal otherwise [22].

Figure 1 presents a pictorial description of the EVSI; the top panel shows the PSA process where the value of the optimal treatment is some monetary value, say $\$C$. To calculate the EVSI, we consider that additional information, denoted by $\boldsymbol{X}$, is going to be collected in a future study, e.g. a clinical trial or an observational study, to reduce the uncertainty in some of the model inputs. If $\boldsymbol{X}$ had been collected, i.e. the future study had already been conducted, then the decision making process would be exactly as we have already described. The information in $\boldsymbol{X}$ would be formally included in the PSA distributions and the value of the optimal decision given this additional information would be either 0, if the average INB is negative, or the average INB itself, equal to $F in the bottom of Figure 1. Comparing $F with $C gives the Value of the Sample Information (VSI).

In general, we can consider the PSA distributions to be the *prior* for $\boldsymbol{\theta}$, which is combined with the data $\boldsymbol{X}$ using Bayes Theorem to compute the posterior distribution for $\boldsymbol{\theta}$. This in turn implies a posterior distribution for the INB.

However, as the future study has not been run (and potentially will never be), we define a *distribution* over the possible, yet unknown study outcomes. The EVSI is then equal to the average VSI over all these possible future datasets. Mathematically, it can be expressed in terms of the INB as

$$\text{EVSI} = E_{\boldsymbol{X}}\big[\max\{0, E_{\boldsymbol{\theta}|\boldsymbol{X}}[INB]\}\big] - \max\{0, E_{\boldsymbol{\theta}}[INB]\}$$

where $E_{\boldsymbol{\theta}|\boldsymbol{X}}[INB]$ is the posterior expectation of the INB for a specific sample $\boldsymbol{X}$. Typically, $\boldsymbol{X}$ will only directly update a small number of model parameters, but in a Bayesian setting we still take expectation with respect to the joint posterior distribution $p(\boldsymbol{\theta}|\boldsymbol{X})$.

To model the possible study outcomes, we specify the sampling distribution for the data, had they been collected. This will depend on some of the model inputs: for example, a binomial distribution models the number of people responding to a treatment conditional on the success rate of the drug. Combining this sampling distribution, conditional on the model inputs, with the PSA distributions for the model inputs gives a distribution over the potential future datasets. This process is represented by the red arrow in Figure 1.

PSA is normally undertaken using a simulation approach [22, 21, 25] where $S$ simulations are taken from the distribution of the model inputs. Each of these simulations is then fed through the economic model to calculate $S$ simulations from the distribution of the INB, denoted $\text{INB}_s$ for $s = 1, \ldots, S$.

To calculate the EVSI, a potential future dataset denoted $\boldsymbol{X}_s$ is then simulated for each PSA simulation $s = 1, \ldots, S$. This uses the sampling distribution for $\boldsymbol{X}$ *conditional* on a subset of the model inputs. Each of these simulated datasets is used to find the posterior for the model inputs and thus the posterior of the INB. In the most general setting, this updating uses Monte Carlo methods, particularly Markov Chain Monte Carlo (MCMC) methods. This means $R$ simulations are taken from the posterior distribution of the model inputs to calculate the average INB.

Consequently, to calculate the EVSI by Monte Carlo requires $S \times R$ simulations, which can be relatively computationally expensive for standard choices of $S$ (normally around 1000) and $R$ (at least 600 [26]).

The method presented in this paper uses a nested Monte Carlo scheme but finds the posterior distribution of the model inputs for a small number of potential future datasets. Specifically, the required number of posterior updates reduces from $S$ to $Q < 100$. This maintains the flexibility of the nested Monte Carlo method whilst drastically reducing its computational cost, irrespective of the model structure.

## 3. The moment matching method

The efficient nested Monte Carlo method [10] is based on "moment matching" and requires several elements. To begin, we must estimate the mean $\mu_\theta$ and variance $\sigma_\theta^2$ of the INB, using the PSA simulations.

### 3.1. Uncertainty due to key parameters

As previously discussed, the sampling distribution of $X$, typically, only depends directly on a small number of the model inputs. For example, within a full health economic model, a clinical trial would focus on the epidemiological parameters and not the economic disease burden. To formalise this, we assume that the model inputs split into two categories, $(\boldsymbol{\phi}, \boldsymbol{\psi})$, where the sampling distribution for $X$ is based solely on the inputs $\boldsymbol{\phi}$ (e.g. the epidemiological parameters) and all the remaining inputs are in $\boldsymbol{\psi}$. The moment matching method requires the distribution of the INB where uncertainty due to the model inputs in $\boldsymbol{\psi}$ has been marginalised out. This is expressed mathematically as

$$\text{INB}_{\boldsymbol{\phi}} = E_{\boldsymbol{\psi}|\boldsymbol{\phi}}[\text{INB}(\boldsymbol{\phi}, \boldsymbol{\psi})].$$

Notice that this quantity is required to calculate an alternative Value of Information measure known as the Expected Value of Partial Perfect Information (EVPPI) [6], which quantifies the economic value of learning the *exact* value of the model inputs $\boldsymbol{\phi}$. As study information cannot give exact information, the EVPPI for $\boldsymbol{\phi}$ is an upper bound for the EVSI. If this upper bound is low then there is no value in a study targeting $\boldsymbol{\phi}$ and so the modeller can discount a trial targeting $\boldsymbol{\phi}$ before determining a sampling distribution for $\boldsymbol{X}$. Therefore, the EVPPI should always be calculated before proceeding to the EVSI [27], which means that the $\text{INB}_{\boldsymbol{\phi}}$ values should already be available. Several methods have been developed to estimate the EVPPI [28, 29, 30] and, more importantly, general purpose software is available to calculate $\text{INB}_{\boldsymbol{\phi}}$ directly from the PSA simulations [31, 32, 33].

### 3.2. The nested posterior variance

The final element required to estimate the EVSI is the *variance* of the posterior INB, across different possible future samples. Specifically, the variance of the posterior INB is inversely proportional to the EVSI. This is because the decision between the two treatment options becomes more certain as the posterior variance of the INB decreases.

To estimate the expected variance of the posterior INB, we use a nested Monte Carlo method. This is the same process as the "gold standard" EVSI estimation method where potential datasets are simulated conditional on values for $\boldsymbol{\phi}$ and used to find posteriors for the model inputs using MCMC. The economic model is then used to calculate posterior distributions of the INB using simulations for the model inputs. The only difference between the two methods is that the moment matching method requires an estimate of the variance of the posterior INB, rather than its mean.

To accurately estimate the EVSI, the variance of the posterior INB must be estimated for a number of potential datasets, say $Q$. Theoretically, $Q$ should be greater than 30, as accurate estimation of the variance depends on the central limit theorem, but, the variance of the posterior INB is sufficiently accurate for values of $Q$ close to 30 [10], provided the future datasets are simulated using the following procedure.

Loosely, we want to "space-out" the simulated future datasets to accurately estimate the EVSI. To achieve this, we find $Q$ equally spaced values of $\boldsymbol{\phi}$ by determining the *quantiles* of the PSA simulations and use these to generate the future datasets. Practically, we proceed by ordering the PSA simulations for each of the model inputs in $\boldsymbol{\phi}$ and selecting the $\frac{q}{Q+1}$th elements in these ordered lists for $q = 1, \ldots, Q$. For example, if $Q = 3$ and the number of PSA simulations $S = 1000$, then each column in $\boldsymbol{\phi}$ should be ordered and the 250th, 500th and 750th elements from each of these lists should be selected.

For each selected row in the ordered $\boldsymbol{\phi}$ list, we simulate *one* potential future sample from the sampling distribution of $\boldsymbol{X}$. These potential future samples are each used to find a posterior distribution for the model inputs using MCMC. A posterior distribution for the INB is calculated for each of the samples and the variance, denoted $\sigma_q^2$ for $q = 1, \ldots, Q$, is calculated.

### 3.3. Calculating the EVSI

To calculate the EVSI, we must combine all these different elements:

1. The mean of the INB from the PSA simulations $\mu_{\boldsymbol{\theta}}$;
2. The variance of the INB from the PSA simulations $\sigma_{\boldsymbol{\theta}}^2$;
3. The $\text{INB}_{\boldsymbol{\phi}}$ values used to calculate the EVPPI;
4. The variance of the $\text{INB}_{\boldsymbol{\phi}}$ values, denoted $\sigma_{\boldsymbol{\phi}}^2$
5. The posterior variance of the INB calculated $Q$ times using nested sampling $\sigma_q^2$ for

$q = 1, \ldots, Q$,

Firstly, we calculate the average posterior variance across the $Q$ nested samples:

$$\sigma_X^2 = \frac{1}{Q} \sum_{q=1}^{Q} \sigma_q^2.$$

Finally, all the above elements are used to rescale the $\text{INB}_{\boldsymbol{\phi}}$ values:

$$\text{INB}^* = \left(\frac{\text{INB}_\phi - \mu_\theta}{\sqrt{\sigma_\phi^2}}\right)\sqrt{\sigma_\theta^2 - \sigma_X^2} + \mu_\theta.$$

This gives $S$ rescaled $\text{INB}_\phi$ values, where $S$ is the number of PSA simulations, which are then used to estimate the EVSI,

$$\widehat{\text{EVSI}} = \frac{1}{S}\sum_{s=1}^{S}\max\{0, \text{INB}_s^*\} - \max\{0, \mu_\theta\}$$

3.4. Toy Example

To clarify the moment matching method, we estimate the EVSI in a very simple setting. The effectiveness measure in this example is the probability of curing a disease. Information from a previous trial implies that the uncertainty in this effectiveness measure for two treatments under consideration can be modelled using the following Beta distributions $\pi_1 \sim Beta(3,4)$ and $\pi_2 \sim Beta(4,3)$. A literature review determines that uncertainty in the incremental cost can be modelled using a Normal distribution $\delta \sim Normal(3,20)$. Therefore, in this example, $\boldsymbol{\theta} = (\pi_1, \pi_2, \delta)$. We then define the INB for this model as

$$\text{INB} = 100(\pi_1 - \pi_2) - \delta \qquad (1)$$

where 100 is the selected willingness to pay. Note that this threshold is used for illustrative purposes as the effectiveness measure is probability of a cure rather than Quality Adjusted Life Years (QALYs).

For this model, we report PSA simulations for $S = 10$ in Table 1. Each parameter is simulated from its distribution and then each simulated row is used to calculate the INB using equation (1). The mean and variance of the INB are calculated as $\mu_\theta = -4.5$ and $\sigma_\theta^2 = 722$.

Using an EVPPI analysis, we found that reducing uncertainty in $\pi_1$, the probability of a cure for treatment 1, has the highest value. Therefore, we design a trial where treatment 1 is given to 20 people and observe how many patients respond. This implies a Binomial sampling distribution for the future study: $\boldsymbol{X} \sim Binomial(20, \pi_1)$.

To calculate the EVSI, we determine the INB conditional on $\phi = \pi_1$. In this example, we have used the EVPPI calculation method developed by Strong et al. 29 to calculate $\text{INB}_\phi$ and the values are given in Table 1. The variance of $\text{INB}_\phi$ is then calculated, $\sigma_\phi^2 = 391$.

The final element required for the moment matching method is the posterior variance estimated using nested sampling. For illustrative purposes, we set $Q = 3$. To begin, we order the observed $\pi_1$ values, which produces the following vector:
$$(0.26, 0.27, 0.30, 0.37, 0.47, 0.50, 0.51, 0.53, 0.59, 0.76).$$

We then select the 2.5th, 5.5th and 7.5th elements in this ordered list. Each of these values is used to simulate a potential future sample from the sampling distribution of $X$. Each future sample generates a posterior for $\pi_1$ which then determines the posterior INB. In this example, 100 simulations from the posterior of $\pi_1$ are combined with 100 simulations of $\pi_2$ and $\delta$ to calculate the posterior INB and its variance. Table 2 demonstrates this process giving the variance of the posterior INB for each sample and the average posterior variance $\sigma_X^2$ as 406.

We can now combine all the elements to rescale the $\text{INB}_\phi$ values,
$$\text{INB}^* = \left(\frac{\text{INB}_\phi - (-4.5)}{\sqrt{391}}\right)\sqrt{722 - 406} + (-4.5).$$
These values are given in Table 1. Finally, the EVSI is estimated using moment matching, as
$$\text{EVSI} = \frac{1}{10}(4 + 10 + 29 + 2) - \max\{0, -4.5\} = 4.6$$

This example is illustrative of the moment matching method; in practice a large PSA simulation size should be used, ideally in excess of 1000.

4. Calculating the EVSI: a new chemotherapy drug

To demonstrate the moment matching method and investigate the optimal value for $Q$, we extend a model developed to compare a new chemotherapy treatment against the standard of care [21]. In this model, the two treatments differ only in the number of side effects experienced, with the new drug aiming to reduce it. This in turn reduces the cost of treating side effects, as they require additional treatment, and increases the patient's quality of life.

There are 14 model inputs related to the cost and QALYs associated with experiencing side effects as well as the probability of experiencing side effects on either treatment. The progression of the side effects is modelled using a four state Markov Model with further details provided in the supplementary material, along with model code.

### 4.1. A trial investigating side effects of Chemotherapy treatment

To gather more information for this model, we designed a trial where both treatments would be given to 150 patients. In the main, this trial will inform the probability of side effects for each treatment. As a secondary outcome, we also study the disease progression of any of the patients experiencing side effects. This directly informs the transition probabilities of the Markov Model for side effects treatment. An EVPPI analysis demonstrated that gaining perfect information about these parameters would explain 90% of the model uncertainty. For simplicity, we assumed complete compliance with the treatment and full follow-up. A full description of the sampling distributions used for the EVSI analysis is given in the supplementary material.

### 4.2. The optimal number of nested samples

To find the optimal number of nested samples, we fix the total number of simulations used to estimate the EVSI. Clearly, as with all simulation based methods, we increase the accuracy of the EVSI estimate by increasing the total number of simulations. However, in fixing the total computational cost we determine the relative importance of increasing $Q$ versus increasing the number of simulations from the posterior distribution of the model inputs.

We use two sizes of nested simulation, 500 000, with a computation time of between 108 and 129 seconds for one EVSI estimate and 5 000, with a computation time of between 9 and 17 seconds. We then considered 8 different values of $Q = 20, 30, \ldots, 100$, where 20 is below the recommended lower limit for $Q$ [10]. As the EVSI is estimated by simulation, it is subject to random variance. Therefore, we calculated the EVSI 200 times for each combination of $Q$ and simulation size. This allows us to calculate the variance of the EVSI estimate for each combination (Figure 2) and the bias (Figure 3). To calculate the bias, we estimated the EVSI using nested simulation with $S = 100\ 000$ and $R = 100\ 000$ with a computational cost of around 60 days.

Figure 2 demonstrates that larger $Q$ values produce more precise EVSI estimates as the variance is smaller. This is true for both 500 000 and 5 000 simulations although the variance seems to plateau earlier for 5 000 implying that increasing $Q$ is less important when the number of posterior simulations is not sufficient to estimate the posterior of $\boldsymbol{\theta}$. Therefore, computational effort should be used to increase $Q$ and not $R$, the number of simulations for each posterior. Clearly, the variance of the estimated EVSI is lower with a greater number of simulation.

Figure 3 demonstrates that larger $Q$ values produce more accurate EVSI estimates. Nevertheless, the moment matching method seems to have a slight upward bias, reducing as $Q$ increases. Notice, that the bias is large for $Q = 20$, confirming that the moment matching method should only be used with $Q > 30$.

In light of this analysis, $Q$ should be chosen in the following manner. Firstly, determine the number of simulations $R$ required to characterise the posterior distribution of $\boldsymbol{\theta}$ and the total number of simulations available to estimate the EVSI. Then, $Q$ should be equal to the total number of simulations divided by $R$. For example, if 50 000 posterior simulations can be used to estimate the EVSI and $R = 1\ 000$, then $Q = 50$ nested simulations should be used.

As a final comment, note that the INB$^*$ values are calculated using the difference between the PSA variance of the INB $\sigma_\theta^2$ and the mean posterior variance $\sigma_X^2$, from Equation (2). Therefore, accurately estimating $\sigma_X^2$ using nested simulations is useless if $\sigma_\theta^2$ is poorly estimated. Thus, the PSA simulations size $S$ must be relatively large. Additionally, the required accuracy of the estimates of $\sigma_X^2$ and $\sigma_\theta^2$ actually depends on the size of the EVSI. For a small EVSI, the posterior variance is close to the prior variance meaning that of $\sigma_X^2 - \sigma_\theta^2$ is small. As both variances are estimated by simulation, the sampling variation could be larger than the "true" difference, leading to inaccurate EVSI estimates. Therefore, the moment matching method should only be used for studies where the EVPPI for $\boldsymbol{\phi}$ is high.

5. Conclusion

In this paper, we have presented an efficient Monte Carlo estimation method for the EVSI. This method significantly reduces the number of nested simulations required to accurately estimate the EVSI. This maintains the flexibility and comprehensibility of the nested Monte Carlo method whist reducing computational time. The moment matching method was presented in theoretical terms alongside an illustrative worked example. Finally, we explored the optimal number of nested samples required to accurately estimate the EVSI. In general, the number of nested samples should be taken as large as possible, whilst maintaining a feasible computational cost and ensuring that the posterior distribution for the model inputs is adequately captured.

Another advantage of this efficient Monte Carlo method is that general purpose software has been developed to perform these EVSI calculations [ 34]. Early results also indicate that this method can be extended to estimate EVSI across different sample sizes with no additional computational cost meaning that the EVSI could be used as a tool for trial design within a realistic time frame.

**Acknowledgments**

Financial support for this study was provided in part by grants from Engineering and Physical Sciences Research Council (EPSRC) [Anna Heath] and Mapi [Dr. Gianluca Baio]. The funding agreement ensured the authors' independence in designing the study, interpreting the data, writing, and publishing the report.